\documentclass[prb,twocolumn,showpacs,superscriptaddress,groupedaddress]{revtex4}
\usepackage[pdftex]{graphicx}  
\usepackage{dcolumn}   
\usepackage{bm}        
\usepackage{amssymb}   
\usepackage{amsmath}   
\usepackage{hyperref}
\usepackage[latin1]{inputenc}
\usepackage{tikz}
\usetikzlibrary{shapes,arrows}

\begin{document}

\title{Exact time-dependent density-functional potentials for strongly correlated tunneling electrons}
\date{\today}
\author{M.\ J.\ P.\ Hodgson} \affiliation{Department of Physics, University of York and European Theoretical Spectroscopy Facility, Heslington, York YO10 5DD, United Kingdom}
\author{J.\ D.\ Ramsden} \affiliation{Department of Physics, University of York and European Theoretical Spectroscopy Facility, Heslington, York YO10 5DD, United Kingdom}
\author{J.\ B.\ J.\ Chapman} \affiliation{Department of Physics, University of York and European Theoretical Spectroscopy Facility, Heslington, York YO10 5DD, United Kingdom}
\author{P.\ Lillystone} \affiliation{Department of Physics, University of York and European Theoretical Spectroscopy Facility, Heslington, York YO10 5DD, United Kingdom}
\author{R.\ W.\ Godby} \affiliation{Department of Physics, University of York and European Theoretical Spectroscopy Facility, Heslington, York YO10 5DD, United Kingdom}

\begin{abstract}
By propagating the many-body Schr\"odinger equation, we determine the exact time-dependent Kohn-Sham potential for a system of strongly correlated electrons which undergo field-induced tunneling. Numerous features are entirely absent from the approximations commonly used in time-dependent density-functional theory. The self-interaction correction is strong and time dependent, owing to electron localization, and prominent dynamic spatial potential steps arise from minima in the charge density, as modified by the Coulomb interaction experienced by the partially tunneled electron.
\end{abstract}

\maketitle

\section{Introduction}

The notable prominence of density-functional theory (DFT) methods in condensed-matter physics stems from the ability to treat the electrons as noninteracting in the Kohn-Sham (KS) approach, \cite{PhysRev.140.A1133} while still, in principle, reproducing the exact charge density of the real interacting system. In practice, the exchange-correlation (xc) part of the KS potential must be approximated, and this is often done on the basis of the local density \cite{PhysRev.140.A1133} or its gradient. \cite{PhysRevLett.77.3865} While these approximations fare surprisingly well in many scenarios, they are known to break down in a large number of important cases. \cite{PhysRevB.69.235411, PhysRevB.73.121403} In time-dependent DFT (TDDFT), the extension of DFT to a system evolving from (typically) its ground state, the time-dependent xc potential depends on the history of the density. However, most time-dependent calculations today use an adiabatic approximation, which assumes dependence only on the instantaneous charge density, \cite{TutorialinDFT} disregarding the system's history and initial state. In this Rapid Communication we study the field-induced tunneling of interacting, strongly correlated electrons through a potential barrier, and identify important features in the exact xc potential that are entirely absent when adiabatic or local approximations are used.

The successful experimental determination of the transport characteristics of quantum junctions \cite{ScienceVol.278, tao2006electron} provides the opportunity to test the predictive power of TDDFT in the presence of electron currents. The usual adiabatic local and semilocal xc approximations prove unreliable. For instance, the current-voltage ($I$-$V$) characteristics of organic molecules, including their conductance, often differ from experiment by orders of magnitude. \cite{PhysRevB.69.235411, DFcalc} Ultra-nonlocal density dependence in the xc potential, including the formation of pronounced spatial steps, has been demonstrated recently for a quasiparticle wavepacket added to a model semiconductor, \cite{PhysRevLett.109.036402} and for a one-dimensional He atom in the presence of a weak oscillatory electric field. \cite{PhysRevLett.109.266404} 

Electron tunneling and reflection are features of all molecular devices, but many-body (MB) aspects of these processes are generally not included in transport calculations. Fundamental studies of tunneling in strongly correlated systems, such as Coulomb blockade, \cite{PhysRevLett.104.236801} have shown the importance of a time-dependent description of electronic correlation. Strong correlation and tunneling both pose particular challenges for the usual approximations in TDDFT which remain to be addressed. To accurately model electron transport there is a need for studies of the xc potential for systems of multiple electrons with strong correlation and quantum tunneling. We calculate the exact xc potential of a one-dimensional interacting model system, intended to inform the development of improved approximate functionals suitable for realistic three-dimensional systems.

\section{Model tunneling system}

To determine the KS potential we adjust the potential experienced by noninteracting electrons such that they reproduce, at all times, the charge and current densities of the interacting system, calculated by exact numerical propagation of the time-dependent many-body Schr\"odinger equation. Our iDEA (interacting dynamic electrons approach) code describes two electrons in one dimension, where our spatial and temporal grid spacings are $\delta x=0.05$ a.u.\ and $\delta t=0.002$ a.u. 

We treat our electrons as spinless, in order to maximize the richness of the correlation for a given computational effort. For example, the two interacting electrons experience the exchange effect (which is not the case for two spinful electrons in an $S=0$ state), and our two Kohn-Sham electrons occupy distinct KS orbitals so that the density is not simply twice the square modulus of the orbital. Both these features, crucial in extended electronic systems, would require a larger number of spinful electrons in order to become apparent.

\begin{figure}[htbp]
  \centering
  \includegraphics[width=1.0\linewidth]{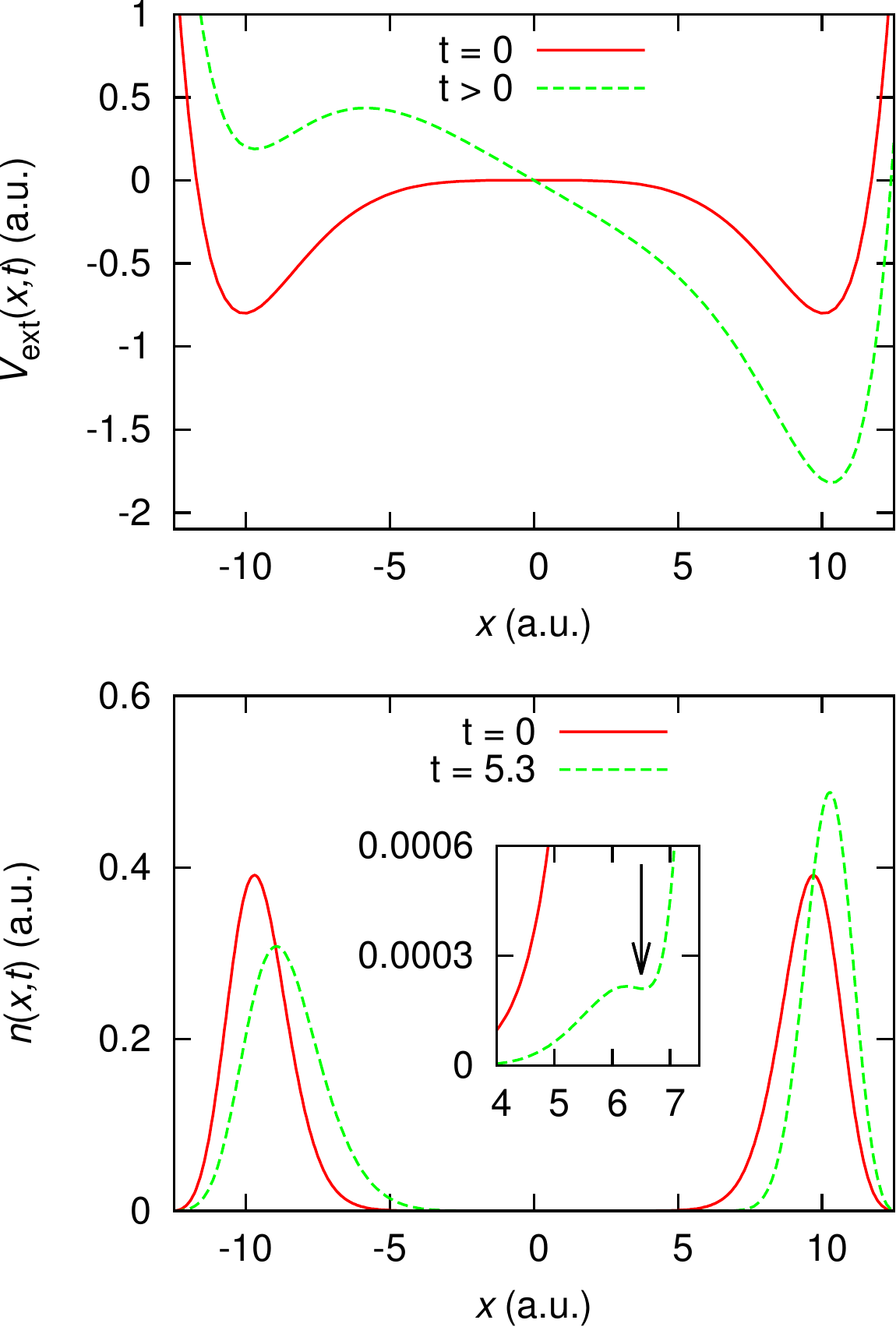}
\caption{(Color online) The unperturbed external potential ($V_{\mathrm{ext}}=\alpha x^{10}-\beta x^{4}$, $t\leq 0$, solid red) and its perturbed counterpart (the uniform field $-\varepsilon x$ added, $t>0$, dashed green) (top figure). The interacting charge density (bottom figure) at $t=0$ and at a later time $t=5.3$ a.u. The inset shows a close-up of the density in the tunneling region; at the later time a minimum appears in the density (arrow) as a result of an interference effect (see text).}
\label{mb_barrier}
\end{figure}

We first describe our system through MB quantum mechanics. Our MB spinless electrons interact via the softened Coulomb term $\left(\left|x_{i}-x_{j}\right|+0.1\right)^{-1}$ in the Hamiltonian, as is appropriate in one dimension. Our confining potential consists of two wells separated by a long flat barrier, $V_{\mathrm{ext}}=\alpha x^{10}-\beta x^4$, where $\alpha=5 \times 10^{-11}$ and $\beta =1.3 \times 10^{-4}$. For $t>0$ a polarizing uniform electric field $- \varepsilon x$, where $\varepsilon=0.1$, is applied (Fig.~\ref{mb_barrier}), driving the electrons to the right.

The interacting ground-state MB wave function, including full correlation effects, is calculated by first evolving an arbitrary, exchange-antisymmetric trial wave function through imaginary time \cite{doi:10.1080/01621459.1949.10483310} in the chosen external potential, including the interaction term. We then apply the electric field and evolve the ground-state wave function through real time. The Crank-Nicolson method \cite{crank-nicol} is used for both imaginary- and real-time propagation. The interacting density is
\begin{equation}
n_0\left(x,t \right) = 2 \int dx_{2}  \left | \Psi \left(x,x_{2},t \right) \right | ^2 .
\label{intden}
\end{equation}

As suggested by the form of the ground-state charge density, the Pauli exclusion principle, for two spinless electrons, tends to localize the electrons in opposite wells, and this strong-correlation effect is enhanced by the Coulomb repulsion. Thus, the barrier region of the system has vanishingly small density (Fig.~\ref{mb_barrier}). 

The initial application of the electric field begins to establish oscillatory motion of the electrons within their respective wells. Prolonged exposure to the \textit{E} field causes the left electron to tunnel through the potential barrier towards the right well, experiencing the Coulomb repulsion. This then allows a current to build in the low-density barrier region. The strength of the \textit{E} field relative to the confining potential means that both electrons acquire considerable kinetic energy within their respective wells. In the right-hand well this results in standing-wave-like ``ripples'' in the density, due to interference between the waves incident on, and reflected from, the right-hand wall. (This phenomenon does not itself rely on the interaction; we have checked that similar interference ripples occur for a single electron in a single perturbed well.) When tunneling begins, the time evolution of the density is affected by the Coulomb repulsion, as the two electrons try to distance themselves from one another within the right-hand well.

\begin{figure}[htbp]
  \centering
  \includegraphics[width=1.0\linewidth]{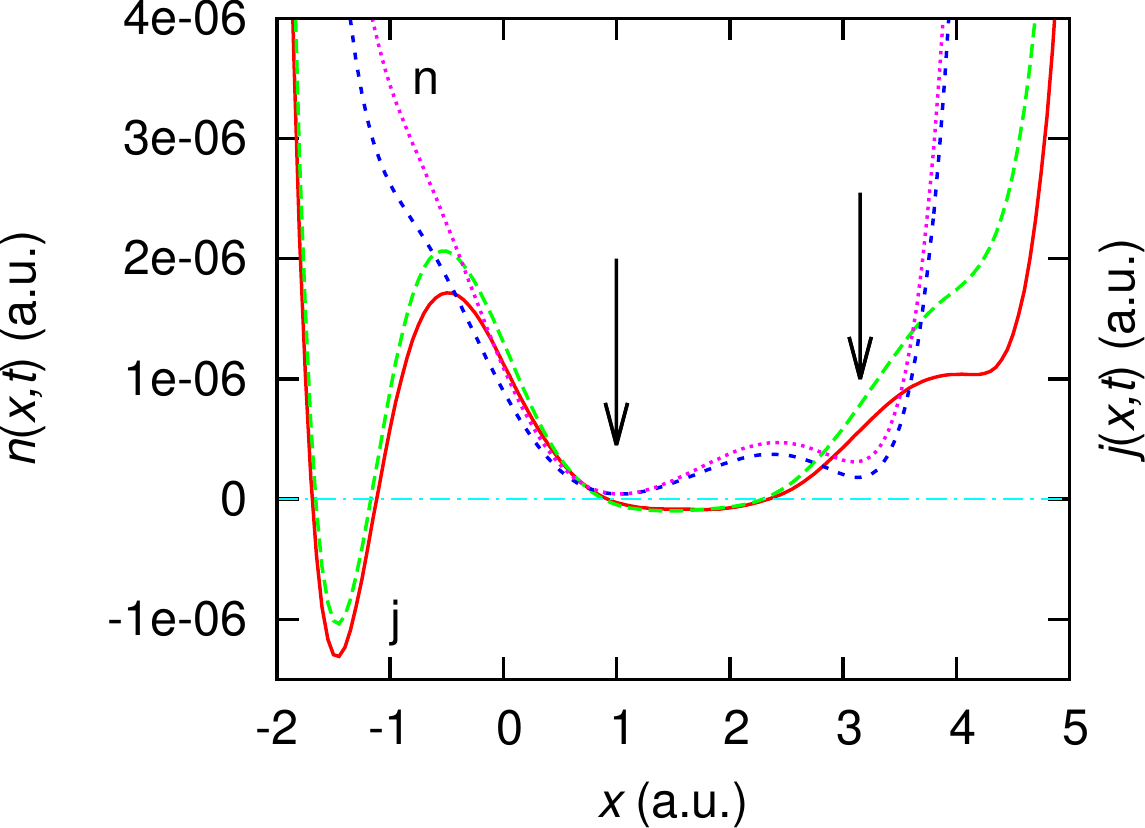}
\caption{(Color online) A zoomed view of the interacting charge density ($n$) in the central region (short dashed blue) and the non interacting charge density (dotted purple), together with the interacting current density ($j$) (solid red) and the non interacting current density (dashed green), at $t = 5.3$ a.u. The regions of particularly high current-to-charge ratio are indicated by arrows; the modification of this ratio by the Coulomb interaction is evident.}
\label{current}
\end{figure}

Figure~\ref{current}, which is a further close-up, shows the charge and current density in the central part of the tunneling region, for $t = 5.3$ a.u., together with the corresponding quantities in the absence of Coulomb interaction. Two effects are evident. First, as the left electron tunnels into the right well, the effect of the Coulomb repulsion is to suppress the current density on the right-hand side of the region shown. Second, the arrows in both Figs.~\ref{mb_barrier} and \ref{current} indicate locations where the ratio of current density to charge density is particularly high.

\section{TDDFT description}

TDDFT describes the evolution of the interacting charge density using an auxiliary non interacting system, with an effective potential $V_{\mathrm{KS}}$ which we now calculate. We use the same numerical methods, where appropriate, for the non interacting and interacting systems, to minimize numerical error. The time-dependent (TD) KS potential allows the dynamics of the density to be completely described by the single-particle KS orbitals which obey
\begin{equation}
i\frac{\partial}{\partial t}\phi^k \left(x, t \right) = \left\{ - \tfrac{1}{2} \frac{\partial^2}{\partial x^2}+V_{\mathrm{KS}} \left(x, t \right) \right\}\phi^k \left(x,t \right)\nonumber ,
\label{TDKSE}
\end{equation}
and yield the electron density through
\begin{equation}
 n \left(x,t \right) = \sum\limits^{2}_{k=1}\vert\phi^k \left(x,t \right)\vert^2 .
\label{NID}
\end{equation}
At $t=0$, before the system becomes dynamic, the ground-state KS potential describes the system. We determine this by iteratively correcting a trial potential using
\begin{equation}
V_{\mathrm{KS}} \rightarrow V_{\mathrm{KS}} + \mu \left[ n \left(x\right)^{p}-n_0 \left(x\right)^{p} \right],
\label{GSNM}
\end{equation}
where $0 < \mu \leq 1$, $n \left(x\right)$ is the ground-state density produced by $V_{\mathrm{KS}}$, and $n_0 \left(x\right)$ is the target ground-state density. This iterative procedure, which builds on that set out in Ref.~\cite{PhysRevA.49.2421}, clearly has the correct fixed point $n=n_0$ for any $p$, and we find especially rapid convergence when $p\approx0.05$ and $\mu=1$. We monitor the convergence using the integrated absolute error in the density, 
\begin{equation}
f_n = \int_{-\infty }^{\infty }  \left | n \left(x \right)-n_0 \left(x \right) \right | dx .
\label{GScost}
\end{equation}
The use of a small value of $p$ focuses the emphasis of the iterative procedure on the low-density regions, where substantial adjustments to the potential are needed, while avoiding oscillatory instabilities arising from unduly large adjustments to the potential in the high-density regions. Reducing $f_{n}$ below $10^{-11}$ a.u. is rapid, requiring around 1500 iterations. We have found this method to be robust, accurate, and fast, for a variety of systems. 

When our system becomes dynamic we implement a variant of the reverse-engineering algorithm of Ramsden and Godby, \cite{PhysRevLett.109.036402} where we iteratively correct a time-dependent KS vector potential, $A_{\mathrm{KS}}$, at each time step. For this method we temporarily switch to an electromagnetic gauge in which the TD KS potential is split into a \textit{static} scalar potential (the ground-state KS potential plus the external applied field) and a time-dependent vector potential. Working in this gauge reduces computational cost by eliminating the need for a spatial integration in every iteration of the algorithm, as well as improving numerical stability. The vector potential is obtained using the iterative procedure
\begin{equation}
A_{\mathrm{KS}} \rightarrow A_{\mathrm{KS}} + \nu \frac{j \left(x,t \right)-j_0 \left(x,t \right)}{n_0\left(x,t \right)} ,
\label{TDRE}
\end{equation}
where (typically) $\nu=0.5$, which causes the non-interacting current density $j$ produced by $A_{\mathrm{KS}}$, to converge towards the interacting current density $j_0$. We calculate $j$ and $j_0$ directly from their respective time-dependent charge densities, via the continuity equation $\partial_t n + \partial_x j = 0$, using a numerical time derivative of the charge density and a numerical integration. This use of the continuity equation guarantees that the two densities $n$ and $n_0$ automatically match at each time, in addition to $j$ and $j_0$, as required.

Having calculated these potentials we transform them to the gauge where the vector potential is zero, so that the TD KS potential is represented completely by a time-dependent scalar potential, as is conventional for finite systems. 

\begin{figure}[htbp]
  \centering
  \includegraphics[width=1.0\linewidth]{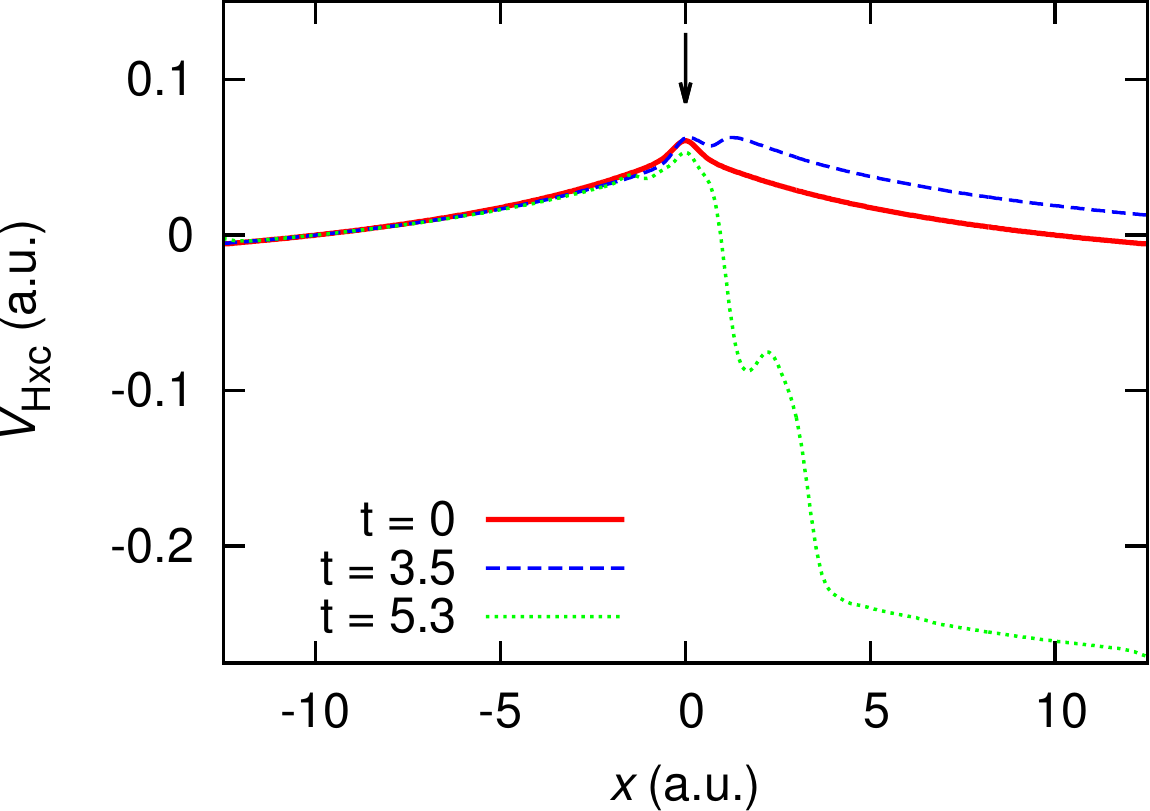}
\caption{(Color online) The Hxc potential in the ground state ($t=0$) and at later times $t=3.5$ and $5.3$ a.u. Steps form and grow as a result of minima in the charge density in the regions of finite current density. The distinctive ``bump'' in the ground-state Hxc potential remains to serve its initial purpose (arrow).}
\label{Hxc}
\end{figure}

The nonlocal density dependence of the xc potential is already apparent in the ground state. Owing to the double-well external potential and the Coulomb repulsion, our electrons begin in a highly localized state, which means that the dominant effect in the ground-state xc potential is the full cancellation of the spurious self-interaction described by the Hartree potential $V_\mathrm{H}$. Self-interaction corrections of this sort are, of course, beyond the capability of the local-density approximation (LDA). The ground-state Hartree-xc (Hxc) potential $V_{\mathrm{Hxc}}=V_{\mathrm{KS}}-V_{\mathrm{ext}}$, because it includes both the self-interaction and its exact cancellation, most clearly displays the \textit{remaining} features of the KS potential. The Hxc potential (Fig.~\ref{Hxc}) shows a highly non-LDA ``bump'' (arrow in Fig.~\ref{Hxc}) between the wells in the region of low charge density, together with oppositely signed xc electric fields within each well \footnote{By removing the self-interaction correction from the xc potential we have calculated the strength of the linear xc electric field in the regions of highest charge density $|\varepsilon_{\mathrm{xc}}| \approx 0.016$ a.u.}, which together push the KS charge density peaks apart to account for the inter-well Coulomb repulsion. Neither of these features is present in local approximations, showing the failure of the LDA in this system.

\begin{figure}[htbp]
  \centering
  \includegraphics[width=1.0\linewidth]{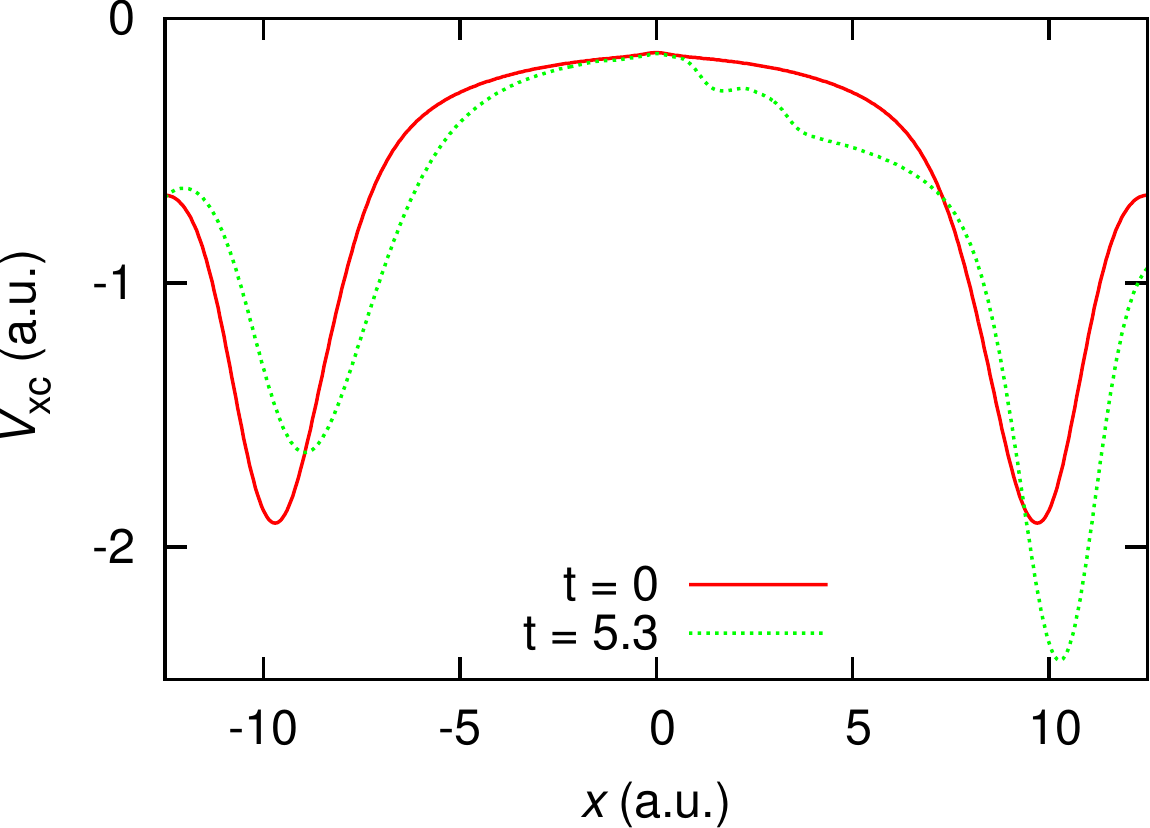}
\caption{(Color online) The xc potential at $t=0$ and $5.3$ a.u. In addition to the features noted in $V_{\mathrm{Hxc}}$, the changing strength of the self-interaction correction as the left electron tunnels into the right well is evident.}
\label{TDXC}
\end{figure}

In the time-dependent regime, the self-interaction continues to be exactly canceled within the Hxc potential. The locally varying corrections to $V_{\mathrm{Hxc}}$ in the region of highest density remain minimal. Figure~\ref{TDXC}, the time-dependent xc potential alone, shows clearly how $V_{\mathrm{xc}}$($=V_{\mathrm{Hxc}}-V_\mathrm{H}$) in the left and right wells provides the necessary self-interaction corrections, changing its form in accordance with the moving and tunneling charge density. 

For the time-dependent Hxc potential (Fig.~\ref{Hxc}) the first feature occurs at the position of the central density minimum, $x_{\mathrm{min}}$ (initially 0). As the left electron tunnels through the barrier, current builds in the central region, and so $x_{\mathrm{min}}$ moves to the right, as already observed in Fig.~\ref{current}. The buildup of current in the vicinity of the density minimum means that the current-to-charge ratio is large. In the KS regime this high ratio is replicated by a prominent time-dependent step in the potential by a positive constant, that later becomes negative due to the change in sign of the current density (Fig.~\ref{Hxc}). 

After enough time has elapsed, further steps form in the right-hand region, as a direct result of the standing-wave effect, associated with the points of high current-to-charge ratio identified above. This ratio, $u\left(x,t\right)=j\left(x,t\right)/n\left(x,t\right)$, the velocity field, can account for the steps observed in the TD KS potential, since it is clear from Eq.~(\ref{TDRE}) that the velocity field and the TD KS potential are closely linked: In particular, a feature in the velocity field will in general be associated with a feature in the TD KS potential. From $u$ it is apparent that when the current-to-charge ratio is very high, i.e. for density minima, there is a peak in the velocity field, which will translate into a TD peak in the KS vector potential. When subsequently transforming each of these peaks to the gauge in which $A_{\mathrm{KS}}=0$ in order to obtain the KS scalar potential, we integrate its time derivative spatially, giving rise to a step, particularly if the peak is narrow, as will generally be the case at density minima. However, the mere presence of a density minimum is not a sufficient condition for a step: Without the effect of the Coulomb repulsion, $j$ and $j_0$ in Eq.~(\ref{TDRE}) can be equal without the need for a correction in the potential. To achieve any feature, the density in the neighborhood of its minimum must be modified by the interaction; in this case this arises primarily from direct interwell Coulomb repulsion. The arrows in Fig.~\ref{current} indicate those density minima that are significantly reduced in value by the Coulomb interaction.

\begin{figure}[htbp]
  \centering
  \includegraphics[width=1.0\linewidth]{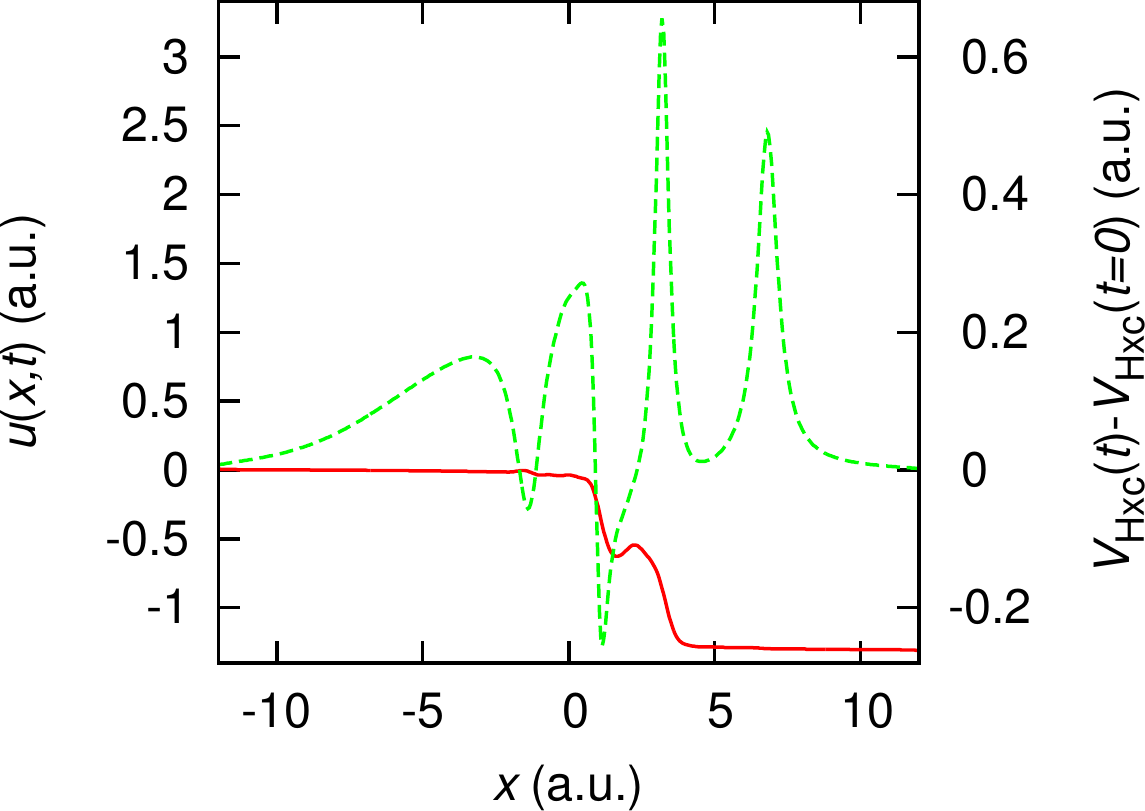}
\caption{(Color online) For $t=5.3$ a.u.\ the velocity field (dashed green, MB and KS coincide), together with the difference between the Hxc potential at $t=5.3$ a.u.\ and the ground-state Hxc potential (solid red). The peaks in the velocity field align with the steps in the potential. The largest features correspond to the density most affected by the Coulomb interaction.}
\label{velocity}
\end{figure}

Figure~\ref{velocity} demonstrates the correlation between peaks in the velocity field and the step functions in the potential. The largest steps are in the central region where the density has been significantly altered by the Coulomb repulsion. Peaks further from the center correspond to smaller steps because the strong initial localization of the electrons reduces the effect of the Coulomb interaction on the velocity field. Our analysis of the origin of steps in the KS potential is quite general; in particular, we have checked that it also explains the TD steps observed in Ref.~\cite{PhysRevLett.109.036402}. Since the two ingredients of the velocity field, the current and charge densities, are always available in a TDDFT calculation, accounting for prominent features of the velocity field should be given strong consideration in the development of improved approximate functionals for use in general time-dependent systems.

\section{Conclusion}

In summary, for these strongly correlated tunneling electrons the exact TD KS potential exhibits density dependence that is ultra-nonlocal in space and nonlocal in time (e.g., through current dependence), aspects that are not present in any local or adiabatic approximations. First, the self-interaction correction is strong, and time dependent as the tunneling occurs. Second, steps occur in the TD KS potential as a consequence of minima in the charge density, combined with the Coulomb repulsion of the electrons on each side of the minimum. Such density minima are a natural consequence of an applied dc electric field, making step functions common and prominent features in the exact TD KS potential. The understanding of the origin of these features in the TD KS potential is crucial information for the development of improved approximate xc functionals that are suited to tunneling and transport systems.

We acknowledge funding from EPSRC.

\bibliography{Bibtex}

\end{document}